\documentclass[aps,prl,twocolumn,superscriptaddress,showpacks]{revtex4-2}

\usepackage{comment}
\usepackage[colorlinks, citecolor=red]{hyperref}
\usepackage{hyperref}
\usepackage[utf8]{inputenc}
\usepackage[T1]{fontenc}    %
\usepackage[english]{babel}
\usepackage[pdftex]{graphicx}
\usepackage{amsmath,amssymb,amsthm,bbm, mathtools} %
\usepackage{mathrsfs}
\usepackage{tikz}   %
\usepackage[normalem]{ulem}

\renewcommand{\vec}[1]{\boldsymbol{#1}}

\begin{document}

\title{Magnetoresistance from decoherence}

\author{Xian-Peng Zhang}

\affiliation{Centre for Quantum Physics, Key Laboratory of Advanced Optoelectronic Quantum Architecture and Measurement (MOE), School of Physics, Beijing Institute of Technology, Beijing, 100081, China}

\author{Yan-Qing Feng}
\email{yq_feng@bitzh.edu.cn}

\affiliation{International Center for Quantum Materials, Beijing Institute of Technology, Zhuhai, 519000, China}





\author{Haiwen Liu}
\affiliation{Center for Advanced Quantum Studies, School of Physics and Astronomy, Beijing Normal University, Beijing 100875, China}

\author{Yugui Yao}
\email{ygyao@bit.edu.cn}
\affiliation{Centre for Quantum Physics, Key Laboratory of Advanced Optoelectronic Quantum Architecture and Measurement (MOE), School of Physics, Beijing Institute of Technology, Beijing, 100081, China}

\begin{abstract}
Microscopic theories of magnetoresistance have traditionally focused on momentum relaxation and the plasma frequency of itinerant electrons. Here, we uncover a distinct mechanism in which magnetoresistance originates from quantum decoherence throughout the whole Fermi sea, specifically the decay of the off-diagonal components of the density matrix. The resulting conductivity, parameterized by two \textit{complex} decoherence times, scales \textit{linearly} with impurity density—markedly contrasting the conventional Drude picture, where conductivity is governed by momentum relaxation of Ferm-surface quasiparticles and is inversely proportional to impurity density. This unconventional scaling provides a direct electrical probe of quantum decoherence, a quantity central to both fundamental studies and emerging nanoscale technologies.  Furthermore, the interplay between the external magnetic field and the exchange field gives rise to rich magnetotransport phenomena, including temperature-drive crossover from positive to negative magnetoresistance and a nonmonotonic temperature dependence with a conductivity maximum reminiscent of the Kondo effect. Our results establish quantum decoherence as a key ingredient in magnetoresistance and our findings should have an unprecedented impact on advancing research and applications involving  magnetoresistance.
\end{abstract}

\maketitle

\textit{Introduction.-}Magnetoresistance (MR)-the change of electrical resistance under an applied magnetic field-serves as a fundamental probe of charge transport and scattering processes in solids~\cite{thomson1857xix,mott1936electrical,mott1936resistance,mott1964electrons,goodings1963electrical,yosida1957anomalous,raquet2002electron,mihai2008electron,zhang2025theoryprl,zhang2023extrinsic,zhang2022microscopic,zhang2019theory}, and underpins a wide range of spintronic and microelectronic technologies~\cite{dieny2020opportunities}. Within the semiclassical framework, MR is governed primarily by momentum relaxation of Fermi-surface quasiparticles, which dictates the decay of the diagonal components of the non-equilibrium density matrix and gives rise to the conventional Drude response of diffusive systems~\cite{drude1902elektronentheorie}. By contrast, in the quantum limit of strong magnetic fields, Landau quantization reorganizes the electronic density of states~\cite{landau1930diamagnetismus}, giving rise--particularly in low-carrier-density semimetals and narrow-gap systems--to a non-saturating MR that is linear in field~\cite{bastin1971quantum,abrikosov1969galvanomagnetic,abrikosov1998quantum,abrikosov2003quantum}. This behaviour has been reported in silver chalcogenides~\cite{xu1997large}, layered semimetals~\cite{abrikosov1999quantum}, and related materials~\cite{friedman2010quantum,wang2021giant,khouri2016linear,novak2015large}, where the quantum limit can be accessed at comparatively modest fields.

In materials with strong spin–orbit coupling and nontrivial band geometry, electronic transport can be profoundly influenced by the Berry curvature of Bloch bands, which acts as an effective magnetic field in momentum space~\cite{xiao2010berry}. This geometric field underlies a variety of anomalous transverse transport phenomena, including the anomalous Hall effect~\cite{nagaosa2010anomalous} and the spin Hall effect~\cite{sinova2015spin}. Microscopically, these phenomena are naturally described by a density matrix that includes off-diagonal components encoding nonequilibrium quantum coherence between Bloch states~\cite{culcer2017interband,culcer2022anomalous}. Despite this progress, two key gaps remain. First, the role of Berry curvature in longitudinal magnetotransport is still poorly understood. Second, existing theories  predominantly focus on momentum relaxation of carrier populations, while largely neglecting the dynamics of quantum coherence and its decay through decoherence processes~\cite{streltsov2017colloquium,zurek2003decoherence}. In systems with strong band-geometry effects, scattering not only relaxes charge-carrying quasiparticles but also induces decoherence between Bloch states~\cite{zhang2026theory}, raising a central question: how this decoherence feeds back into longitudinal magnetotransport constitutes a central unresolved problem?

\begin{figure}[t]
\begin{center}
\includegraphics[width=0.48\textwidth]{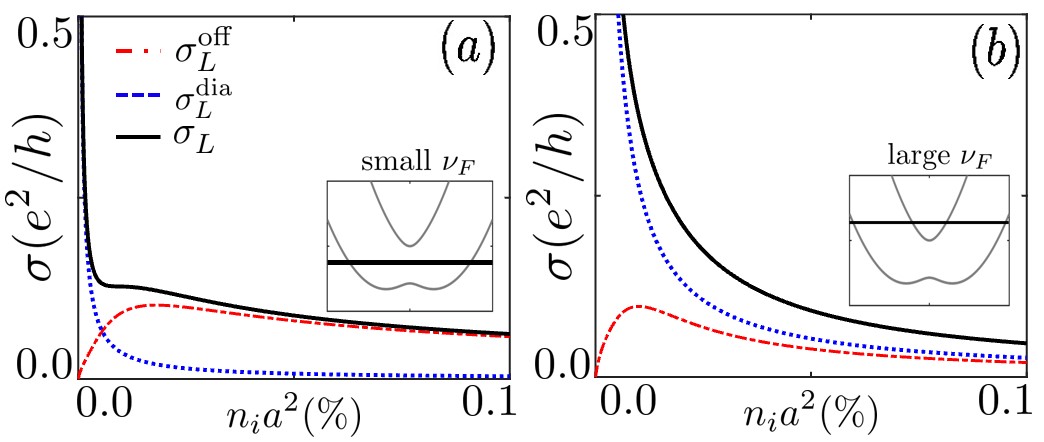}  
\end{center}
\caption{(a) The $n_{\text{i}}$ dependence of $\sigma^{\text{dia}}_{L}$ [Eq. \eqref{fvafkavk}], $\sigma^{\text{off}}_{L}=\sigma^{\Vert}_{L}+\sigma^{\perp}_{L}$ [Eqs. \eqref{oLov} and \eqref{oLop}],  and $\sigma^{}_{L}=\sigma^{\text{dia}}_{L}+\sigma^{\text{off}}_{L}$ where $\epsilon_F=0$ meV the Fermi energy only cuts $\eta=-$ band. Panel (b)  corresponds to $\epsilon_F=0.5$ meV where the Fermi energy cuts both $\eta=-$ and $\eta=+$ bands. Other parameters: $v_R/a=1$ meV, and $\epsilon_L=0.2$ meV, $U/a^2=0.8$ eV, and $a=3.5$ nm.}
\label{FIG3}
\end{figure}

In this Letter, we identify a distinct mechanism of MR arising from quantum decoherence in Berry-curvature-dominated systems. In contrast to conventional Drude MR and quantum linear MR, which originate from quantum population relaxation governed by momentum scattering, the MR reported here stems from the relaxation of quantum coherence induced by randomly distributed impurity densities. We show that this decoherence process generates a longitudinal magnetoresistance controlled by Berry curvature and band geometry. Remarkably, the resulting conductivity is \textit{directly} proportional to the  impurity density through the decoherence time, in sharp contrast to the conventional Drude behavior where conductivity is inversely proportional to impurity density through the momentum relaxation time (Fig. \ref{FIG3}). Our results therefore reveal a previously overlooked origin of magnetotransport beyond the standard population-relaxation paradigm, and suggest that magnetotransport measurements may provide a direct electrical probe of quantum decoherence, a key quantity for both fundamental studies and emerging quantum  technologies.

\textit{Model.-}To study the MR arising from decoherence effects, we employ a non-equilibrium quantum kinetic framework for itinerant electrons described by the total Hamiltonian  $\hat{H}=\hat{H}_{e}+\hat{H}_{E}+\hat{V}$~\cite{zhang2026theory}. The electronic contribution is written as a sum over single-particle operators $\hat{H}_e=\sum_l \hat{\mathcal{H}}^e_0(\vec{r}_l)$, while the interaction with an external electric field is given by $\hat{H}_{E}=-\sum_l e\vec{E}\cdot \vec{r}_l$, and impurity scattering is modeled by $\hat{V}=\sum_{jl}U_0\delta(\vec{r}_l-\vec{R}_j)$. Here $e<0$ denotes the electron charge. The single-particle Hamiltonian $\hat{\mathcal{H}}^e_0(\vec{r}_l)$ incorporates kinetic energy, Rashba spin–orbit coupling, and spin splitting due to both exchange and external magnetic fields  $\hat{\mathcal{H}}^e_0(\vec{r}_l)=\frac{\hbar^2k^2_l}{2m}- v_{R} k_l^ys^x+v_{R} k_l^xs^y+(\epsilon_B+\epsilon_M)\hat{s}^z_l$,  where $\vec{k}_l=-i\nabla{\vec{r}_l}$ is the canonical momentum operator,  $v_R$ characterizes the Rashba coupling strength, and $\hat{\vec{s}}_l=(\hat{s}_l^x,\hat{s}_l^y,\hat{s}_l^z)$ are Pauli matrices acting on the $l$th electron. The total spin splitting is parameterized by $\epsilon_L=\epsilon_B+\epsilon_M$, where $\epsilon_M$ originates from exchange coupling to localized moments and  $\epsilon_B=g\mu_B B$ corresponds to an external magnetic field applied along the $z$ direction with $g$ the electron g-factor and $\mu_B$ the Bohr magneton, for a magnetic field $\vec{B}\parallel \hat{z}$. The energy spectrum of  $\hat{\mathcal{H}}^e_0$ consists of two isotropic branches labeled by $\eta=\pm$, with dispersion $\epsilon_{\vec{k}\eta}=\epsilon_{k}+\eta\mathcal{E}_{\vec{k}}$, where $\epsilon_{k}=\hbar^2k^2/2m$ and $\mathcal{E}_{k}=\sqrt{ v^2_{R}k^2+\epsilon_{L}^2}$. The corresponding eigenstates (see Ref.~\cite{zhang2026theory}) take the form $\vert \vec{k}+ \rangle=\begin{bmatrix}
        \cos\frac{\Theta_{k}}{2}e^{-i\theta_{\vec{k}}}&
        +i\sin\frac{\Theta_{k}}{2}
    \end{bmatrix}^T$ and $\vert \vec{k}- \rangle=\begin{bmatrix}
        \sin\frac{\Theta_{k}}{2}e^{-i\theta_{\vec{k}}} &
        -i\cos\frac{\Theta_{k}}{2}
    \end{bmatrix}^T$, 
with $\cos\Theta_{k}=\epsilon_L/\mathcal{E}_{\vec{k}}$, $\sin\Theta_{k}=v_{R}k/\mathcal{E}_{\vec{k}}$, and $\theta_{\vec{k}}=\text{angle}\left(\frac{k_x+ik_y}{k}\right)$.

\textit{Theory.-}In a dirty material, impurity-induced electron scattering becomes appreciable. Here, we study how the quantum decoherence -- the relaxation of the off-diagonal density matrix generates magnetoresistance from a microscopic perspective. Following the method of Ref.~\cite{zhang2026theory}, we attain the solution of the off-diagonal density matrix 
\begin{align} \label{ansatz}
    \delta\varrho^{\bar{\eta}\eta}_{\vec{k}}=\delta\varrho^{\bar{\eta}\eta}_{\vec{k},\Vert}+\delta\varrho^{\bar{\eta}\eta}_{\vec{k},\perp},
\end{align}
with
\begin{align} \label{fvfkvmkdf4}
\delta\varrho^{\bar{\eta}\eta}_{\vec{k},\Vert/\perp}=-\frac{e}{\hbar}\tau^{\bar{\eta}\eta}_{k,\Vert/\perp}(f_{k\bar{\eta}}- f_{k\eta})\vec{\mathcal{R}}^{\bar{\eta}\eta}_{\vec{k}}\cdot \vec{E}_{\Vert/\perp},
\end{align}
where $\vec{E}_{\Vert}=\vec{E}$ and $\vec{E}_{\perp}=\vec{E}\times \hat{z}$.  The complex normal and anomalous decoherence time read
\begin{align} \label{sgfbf1}
    \frac{\tau^{\bar{\eta}\eta}_{k,\Vert}}{\hbar}=\frac{(\epsilon_{k\bar{\eta}}-\epsilon_{k\eta})-i\Gamma_{k}}{[(\epsilon_{k\bar{\eta}}-\epsilon_{k\eta})-i\Gamma_{k}]^2+(\Gamma^a_{k})^2},
\end{align}
\begin{align} \label{sgfbf2}
    \frac{\tau^{\bar{\eta}\eta}_{k,\perp}}{\hbar}=\frac{\eta\Gamma^a_{k}}{[(\epsilon_{k\bar{\eta}}-\epsilon_{k\eta})-i\Gamma_{k}]^2+(\Gamma^a_{k})^2}.
\end{align}
The ordinary scattering causes a normal decoherence rate $\Gamma_{k}/\hbar=\frac{1}{8}\left(1+\frac{\epsilon^2_L}{\mathcal{E}^2_k}\right)\left(\frac{1}{\tau^{0}_{k+}}+\frac{1}{\tau^{0}_{k-}}\right)$, and the anomalous scattering is quantified by an anomalous decoherence rate $\Gamma^a_{k}/\hbar= \frac{\epsilon_L}{4\mathcal{E}_k}  \left(\frac{1}{\tau^{0}_{k+}}+\frac{1}{\tau^{0}_{k-}}\right) $ where $1/\tau^0_{k\eta}=(2\pi/\hbar)n_{\text{i}}\nu_{\eta}(\epsilon_{k\eta})U^{2}$~\cite{zhang2026theory}. 
The density of state for each energy band is given by $\nu_{\eta}(\epsilon)=\left.\frac{1}{h}\frac{\kappa}{\left\vert v_{\kappa\eta}\right\vert}\right\vert_{\epsilon=\epsilon_{\kappa\eta}}$  with $v_{k\eta}=\frac{\hbar k}{m}+\eta\frac{v_R}{\hbar}\sin\Theta_{k}$. Our solution \eqref{ansatz} recovers  the perfect coherence case with  $\tau^{\bar{\eta}\eta}_{k,\Vert}=\hbar/(\epsilon_{k\bar{\eta}}- \epsilon_{k\eta})$ and $\tau^{\bar{\eta}\eta}_{k,\perp}=0$~\cite{culcer2017interband}, where the nonequilibrium off-diagonal component of density matrix, in steady state, is  proportional to the Berry connection $\vec{\mathcal{R}}^{\bar{\eta}\eta}_{\vec{k}}$, i.e., $\delta\varrho^{\bar{\eta}\eta}_{\vec{k}}=-\frac{ f_{k\bar{\eta}}- f_{k\eta}}{\epsilon_{k\bar{\eta}}- \epsilon_{k\eta}}\vec{\mathcal{R}}^{\bar{\eta}\eta}_{\vec{k}}\cdot e\vec{E}$~\cite{culcer2017interband,sekine2017quantum,atencia2022semiclassical}. When an electron occupies coherent superpositions of $\eta=+$ and $\eta=-$ states, quantum decoherence of the entire Fermi sea enables fruitful quantum transport phenomena, which can be described by the charge current defined as follows
\begin{align} \label{fvdjvjdj}
      \vec{J}^{\alpha}_{}=-\frac{e^2}{\hbar}\sum_{\vec{\vec{k}}\eta}i(\epsilon_{k\eta}-\epsilon_{k\bar{\eta}})\frac{\tau^{\bar{\eta}\eta}_{k,\alpha}}{\hbar}(f_{k\bar{\eta}}- f_{k\eta})\vec{\mathcal{R}}^{\eta\bar{\eta}}_{\vec{k}}\vec{\mathcal{R}}^{\bar{\eta}\eta}_{\vec{k}}\cdot \vec{E}_{\alpha},
\end{align}
where $\alpha=\Vert,\perp$ and $f_{k\eta}=1/(e^{\beta\epsilon_{k\eta}}+1)$ is the Fermi–Dirac distribution with inverse temperature $\beta=1/(k_BT)$ and energy spectrum $\epsilon_{\vec{k}\eta}$. However, there exists a disappearance of MR for the perfect coherence, as the anomalous velocity is perpendicular to external electric field. Bellow, we identify a distinct and ubiquitous mechanism of MR from the impurity-induced decoherence.

\textit{MR from decoherence.-}In the presence of impurity-induced decoherence, the off-diagonal component of the density matrix $\delta\varrho^{\bar{\eta}\eta}_{\vec{k}}$  also generates longitudinal spin-polarized charge current, whose  charge conductivity read $\sigma^{\text{off}}_{L}=\sigma^{\Vert}_{L}+\sigma^{\perp}_{L}$ with
\begin{align} \label{fdvkfk}
    \sigma^{\Vert}_{L}&=-\frac{e^2}{2 h} \sum_{\eta}\eta\int^{\infty}_0  dkf_{k\eta}\frac{v^2_Rk}{\mathcal{E}_k\hbar}\left(1+\frac{\epsilon^2_L}{\mathcal{E}^2_k}\right) \text{Im}(\tau^{+-}_{k,\Vert}),
\end{align}
\begin{align} \label{fdvkfpk}
    \sigma^{\perp}_{L}&=-\frac{e^2}{h} \sum_{\eta}\eta\int^{\infty}_0  dkf_{k\eta}\frac{ v^2_Rk\epsilon_L}{\mathcal{E}^2_k\hbar}\text{Re}(\tau^{+-}_{k,\perp}).
\end{align}  
Importantly, in the high-temperature regime one finds
$\sum_{\eta}\eta f_{k\eta}\simeq-\mathcal{E}_{k}/(2k_BT)$.
Consequently, the decoherence-induced conductance scales inversely with temperature.  Such an unconventional scaling that associates with the coherent superposition of valence and conductance bands throughout the entire Fermi sea, bears a strong resemblance to strange-metal behavior, most prominently observed in cuprate superconductors above their critical temperature~\cite{varma2020linear,cooper2009anomalous,jin2011link,licciardello2019electrical,bruin2013similarity,legros2019universal}. However, disentangling the decoherence contribution  from the conventional Drude response of Fermi-surface quasiparticles is experimentally challenging, as both mechanisms typically coexist in realistic systems. A promising strategy is to focus on Berry-curvature-dominated materials with a small or vanishing density of states at the Fermi level, where Drude transport is intrinsically suppressed~\cite{zhang2026quantum}.

\begin{figure}[t!]
\begin{center}
\includegraphics[width=0.48\textwidth]{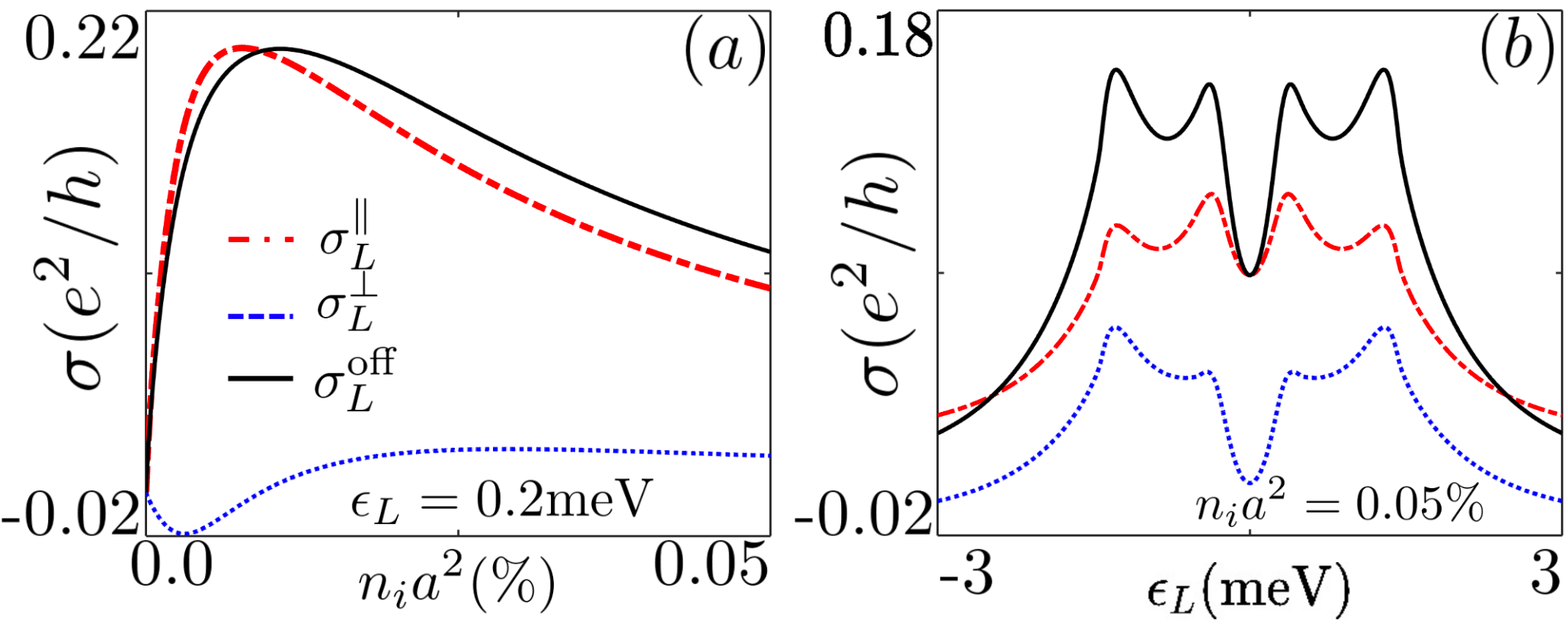} 
\end{center}
\caption{(a) The $n_{\text{i}}$ dependence of  dependence of $\sigma^{\Vert}_{L}$ [Eqs. \eqref{oLov}],
$\sigma^{\perp}_{L}$ [Eq. \eqref{oLop}], and $\sigma^{\text{off}}_{L}=\sigma^{\Vert}_{L}+\sigma^{\perp}_{L}$. Panel (b) plots the corresponding $\epsilon_L$ dependence. Here, $v_R/a=3$ meV and $\epsilon_F=0.5$meV. Other parameters are the same as Fig. \ref{FIG3}. }
\label{Long}
\end{figure}

Next we focus on zero-temperature case, where  Eqs.~\eqref{fdvkfk} and \eqref{fdvkfpk} reduce to
\begin{align}  \label{oLov}
    \sigma^{\Vert}_{L}&=-\frac{e^2}{h} \frac{v_R}{2\hbar}\int^{k^{+}_F}_{k^{-}_F} dk\left(1+\frac{\epsilon^2_L}{\mathcal{E}^2_k}\right)\frac{v_Rk}{\mathcal{E}_k}\text{Im} (\tau^{+-}_{k,\Vert}),
\end{align}
\begin{align}  \label{oLop}
    \sigma^{\perp}_{L}&=-\frac{e^2}{h} \frac{v_R}{\hbar} \int^{k^{+}_F}_{k^{-}_F}  dk \frac{v_Rk}{\mathcal{E}_k}\frac{\epsilon_L}{\mathcal{E}_k} \text{Re} (\tau^{+-}_{k,\perp}).
\end{align}
Quantitative results from Eqs.~\eqref{oLov} and \eqref{oLop} for different values of $n_{i}$ are shown in Figs.~\ref{Long}(a) and (b). Both $\sigma^{\Vert}_{L}$ and $\sigma^{\perp}_{L}$
vanish completely in the absence of quantum decoherence ($n_{\text{i}}=0$). Note that $\sigma^{\Vert}_{L}$ is governed by $\tau^{+-}_{k,\Vert}$ [Eq.~\eqref{sgfbf1}], whose imaginary part  reduces to $\text{Im}(\tau^{+-}_{k,\Vert}/\hbar)\simeq \Gamma_k/(4\mathcal{E}^2_k)\propto n_{\text{i}}$ in the dilute limit, where  Eq.~\eqref{oLov} becomes 
\begin{align} \label{favdfv}
    \sigma^{\Vert}_{L}&=-
\frac{e^2\hbar}{h\tau_0}\sum_{\eta}
\eta\left[ \frac{1}{32\mathcal{E}_{k^{\eta}_F}}
\left(\frac{\epsilon_L^4}{\epsilon^4_R}+\frac{2\epsilon_L^2}{\epsilon^2_R}\right)
+\frac{1}{96\mathcal{E}_{k^{\eta}_F}}\frac{\epsilon_L^4}{\epsilon^2_R\mathcal{E}^2_{k^{\eta}_F}}\right.
\notag\\
& \left.+\frac{1}{64\epsilon_R}
  \left(1 +\frac{\epsilon_L^2}{\epsilon^2_R}\right)^2
  \log\left(\frac{\mathcal{E}_{k^{\eta}_F}-\epsilon_R}{\mathcal{E}_{k^{\eta}_F}+\epsilon_R}\right)  \right],
\end{align}
where the impurity scattering is described by  $\tau^{-1}_0=\frac{n_{\text{i}}mU^2}{\hbar^3}$ and the Rashba spin-orbit coupling is parameterized by $\epsilon_R=mv^2_R/\hbar^2$. 
Thus, the decoherence-induced conductivity grows \textit{linearly} with $n_{\text{i}}$ (i.e., $\sigma^{\Vert}_{L}\propto n_{\text{i}}$),  as shown in Fig. \ref{Long}(a), indicating that small amount of impurities can produce observable  MR from decoherence. Increasing the impurity density enhances $\Gamma_{k}$ ($\propto n_{\text{i}}$), leading to a larger conductivity, contrasting with conventional scenarios where conductivity typically decreases with $n_{\text{i}}$~\cite{drude1902elektronentheorie} (see more details below). However, this MR is suppressed at high impurity density, as coherence dissipates. 
Besides, $\sigma^{\perp}_{L}$ is governed by the anomalous decoherence time $\tau^{+-}_{k,\perp}$ [Eq.~\eqref{sgfbf2}], whose real part  reduces to $\text{Re}(\tau^{+-}_{k,\perp}/\hbar)\simeq -\Gamma^a_k/(4\mathcal{E}^2_k)\propto n_{\text{i}}$ for small impurity density in which  Eq.~\eqref{oLop} reduces to 
\begin{align} \label{favdfp}
    \sigma^{\perp}_{L}\simeq 
\frac{e^2}{h} \frac{\hbar\epsilon_L^2}{16\tau_0\epsilon^2_R}
\sum_{\eta}\eta\left[
\frac{2}{\mathcal{E}_{k^{\eta}_F}}
+\frac{1}{\epsilon_R}\log\left(\frac{\mathcal{E}_{k^{\eta}_F}-\epsilon_R}{\mathcal{E}_{k^{\eta}_F}+\epsilon_R}\right)\right].
\end{align}
Again, the decoherence-induced conductivity scales with $n_{\text{i}}$ (i.e., $\sigma^{\perp}_{L}\propto n_{\text{i}}$), as shown in Fig. \ref{Long}(a). 
Since $\Gamma^a_{k}\propto \epsilon_L$, $\sigma^{\perp}_{L}$ vanishes as $\left\vert \epsilon_L \right\vert\rightarrow0$ and increases with small $\left\vert \epsilon_L \right\vert$, as plotted by the blue curve in Fig. \ref{Long}(b). Notably, the logarithmic term  in Eqs. \eqref{favdfv} and \eqref{favdfp} reveals resonant structure at $(v_Rk^{\eta}_F)^2+\epsilon^2_L=\epsilon^2_R$. This is the Rashba–exchange matching condition, and generates four resonant peaks in Fig.~\ref{Long}(b). Together, $\sigma^{\Vert}_{L}$ and $\sigma^{\perp}_{L}$ demonstrate the essential role of decoherence in modulating MR phenomena.

\textit{Comparison with MR from momentum relaxation.-}Finally, we clarify how the decoherence-induced conductivity can be experimentally distinguished from the conventional Drude contribution. In the Drude picture~\cite{drude1902elektronentheorie}, the longitudinal conductivity  arises from momentum relaxation of the Fermi-surface quasiparticles--the lifetime of the diagonal component of the density matrix
\begin{align} \label{fvafkavk}
   \sigma^{\text{dia}}_{L}=\frac{e^2}{2h}\sum_{\eta} \frac{4\mathcal{E}^2_{k^{\eta}_F}}{4\epsilon^2_L+(v_Rk^{\eta}_F)^2}k^{\eta}_F   v_{k^{\eta}_F\eta} \tau^{0}_{k^{\eta}_F\eta}.
\end{align}
Since the transport lifetime scales as $\tau^{0}_{k^{\eta}_F\eta}\propto 1/n_{\mathrm{i}}$, the Drude conductivity follows $\sigma^{\text{dia}}_{L}\propto1/n_{\mathrm{i}}$. In sharp contrast, the decoherence-induced contribution is governed by the decoherence time and scales linearly with impurity density, $\sigma^{\text{off}}_{L}\propto n_{\mathrm{i}}$ in the dilute limit [see Eqs. \eqref{favdfv} and \eqref{favdfp}]. This linear-in-$n_{\mathrm{i}}$ scaling is consistent with Green’s-function approaches to quantum magnetoresistance developed by Bastin~\cite{bastin1971quantum} and Abrikosov~\cite{abrikosov1969galvanomagnetic,abrikosov1998quantum,abrikosov2003quantum}, although those works predict non-saturating behavior with increasing magnetic field.   Consequently, the two mechanisms exhibit opposite impurity-density dependences: $\sigma^{\mathrm{dia}}_{L}\propto 1/n_{\mathrm{i}}$ and $\sigma^{\mathrm{off}}_{L}\propto n_{\mathrm{i}}$. This contrast provides a clear experimental diagnostic to disentangle the two contributions. As shown in Figs.~\ref{FIG3} and \ref{Long}, the decoherence-induced components ($\sigma_L^{\parallel}$, $\sigma_L^{\perp}$, and $\sigma^{\mathrm{off}}_L$) increase linearly with $n_{\mathrm{i}}$, whereas the Drude contribution decreases with increasing disorder strength, leading to a competition between the two transport channels. Physically, this distinct behavior reflects their different microscopic origins. In the Drude picture, transport is dominated by carriers at the Fermi surface [see Eq.~\eqref{fvafkavk}]. By contrast, decoherence-driven transport involves coherent quantum processes that include electronic states throughout the Fermi sea [see Eqs.~(\ref{fdvkfk}-\ref{oLop})]. Such contributions become particularly important in Berry-curvature-dominated systems with a small or vanishing density of states at the Fermi energy $\nu_F$, where conventional Drude transport is strongly suppressed~\cite{zhang2026quantum}. When the Fermi energy intersects only the $\eta=-$ band, corresponding to a smaller $\nu_F$, the total conductivity is dominated by the decoherence mechanism, as shown in Fig.~\ref{FIG3}(a). In contrast, when both $\eta=-$ and $\eta=+$ bands are occupied, yielding a larger $\nu_F$, the decoherence and Drude contributions become comparable [see Fig.~\ref{FIG3}(b)].

\textit{Failure of scaling analysis.-}The conventional understanding of experiments of anomalous and spin Hall effects is based on three canonical contributions: skew scattering ($\sigma_{H}^{\mathrm{sk}}\propto\sigma_{L}$), side jump ($\sigma_{H}^{\mathrm{sj}}\sim \mathrm{const}$), and intrinsic mechanisms ($\sigma_{H}^{\mathrm{int}}\sim \mathrm{const}$)~\cite{nagaosa2010anomalous,sinova2015spin}. Within this framework, a scaling relation $\rho_{xy}=\alpha\rho_{L}+\beta\rho_{L,0}^2+b\rho_{L}^2$ is obtained~\cite{onoda2006intrinsic,onoda2007quantum,tian2009proper,smit1955spontaneous,smit1958spontaneous,berger1970side,karplus1954hall,jungwirth2002anomalous,nagaosa2010anomalous}. This analysis relies on two key assumptions: (i) the transverse response is small ($\sigma_{H}\ll\sigma_{L}$), such that $\rho_{xy}\simeq-\sigma_{xy}/\sigma_{xx}^2$ and $\sigma_{xx}\simeq 1/\rho_{xx}$; and (ii) the skew-scattering contribution scales proportionally with the longitudinal resistivity. In Berry-curvature-dominated systems, however, neither assumption generally holds, as the longitudinal conductivity can acquire a decoherence origin that can be larger than Drude contribution [Fig.~\ref{FIG3}(a)]. As a result, the conventional scaling breaks down. It is therefore essential to first determine whether the longitudinal response is governed by Drude transport or decoherence mechanisms before applying any scaling analysis.

\begin{figure}[t]
\begin{center}
\includegraphics[width=0.48\textwidth]{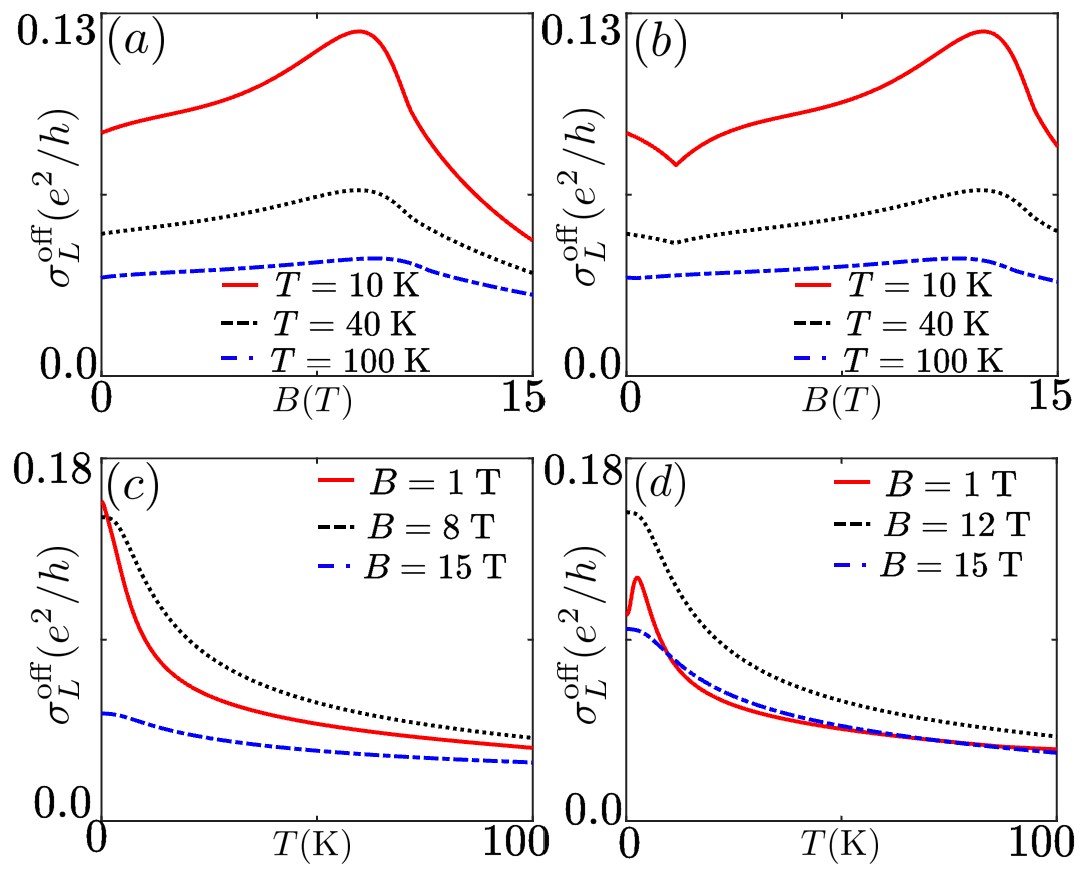}   
\end{center}
\caption{(a-b) $B$ dependence of $\sigma^{\text{off}}_{L}$ for (a)   $n_S\mathcal{J}_{sd}=+0.2$ meV and (b)  $n_S\mathcal{J}_{sd}=-0.2$ meV. Panels (c) and (d) plot the corresponding $T$ dependence. Here, $n_ia^2=0.02\%$, $S=1$, $T_c=100$ K,  and other parameters are the same as Fig. \ref{Long}. }
\label{FIG2q}
\end{figure}

\textit{Temperature and field dependence.-}In realistic experiments, experimental measurements of magnetic field dependence typically deviate from the $\epsilon_L$ dependence plotted in Fig.~\ref{Long} (a) and (b), as  the exchange field $\epsilon_M$  also relies on the temperature and external magnetic field through the spin expectation $\langle S^{}_{\Vert }\rangle$ as follows~\cite{zhang2024microscopic,zhang2025open}
\begin{align} \label{fvdkvkakf}
   \epsilon_{M}=-n^{}_{\mathrm{S}}\mathcal{J}_{sd}\langle S^{}_{\Vert }\rangle,
\end{align}
where $n_{\text{S}}$ is the density of local moments, and 
positive (negative) $\mathcal{J}_{sd}$ corresponds to ferromagnetic (antiferromagnetic) spin exchange coupling, and $S_{\Vert }$ is the spin component in the direction of the magnetization, which is assumed to be collinear to $\vec{B}$. 
Then, the conductivity from normal and anomalous decoherence rate, at finite temperature, are given by 
\begin{align} 
    \sigma^{\Vert}_{L}&=-\frac{e^2}{2 h} \sum_{\eta}\eta\int^{\infty}_0  dkf_{k\eta}\frac{v^2_Rk}{\mathcal{E}_k\hbar}\left(1+\frac{\epsilon^2_L}{\mathcal{E}^2_k}\right) \text{Im}\left(\tau^{+-}_{k,\Vert}\right),
\end{align}
\begin{align} 
    \sigma^{\perp}_{L}&=-\frac{e^2}{h} \sum_{\eta}\eta\int^{\infty}_0  dkf_{k\eta}\frac{ v^2_Rk\epsilon_L}{\mathcal{E}^2_k\hbar}\text{Re}\left(\tau^{+-}_{k,\perp}\right),
\end{align} 
which, at zero temperature, reduce to Eqs.~\eqref{oLov} and \eqref{oLop}, respectively. The dependence of conductivities on $T$ arises from $\langle S_{\Vert}\rangle $ via $
\epsilon_{M}$, while that of $B$ has an extra
channel - the spin precession frequency $
\epsilon_{B}(\propto B)$.  Figures~\ref{FIG2q}(a) and (b) show the magnetic-field dependence of $\sigma^{\parallel}_{L}$ and $\sigma^{\perp}_{L}$ at several temperatures. Notably, for negative $\mathcal{J}_{sd}$, a crossover from positive to negative magnetoresistance emerges upon lowering the temperature [Fig.~\ref{FIG2q}(b)]. The exchange field~\eqref{fvdkvkakf} is proportional to $\mathcal{J}_{sd}\langle S_{\parallel}\rangle$. In the low-temperature, i.e., strongly magnetized regime [red curve of Fig.~\ref{FIG2q}(b)], this field saturates at a negative value and becomes essentially independent of the external field. As a result, increasing the external field initially reduces the net Zeeman splitting, leading to a suppression of the conductivity. The interplay between the external field and the exchange field further gives rise to a nonmonotonic temperature dependence, with a maximum in the conductivity reminiscent of the Kondo effect [Fig.~\ref{FIG2q}(d)].

\textit{Conclusion.-}In this Letter, we have identified a distinct mechanism of magnetoresistance rooted in quantum decoherence in Berry-curvature-dominated systems. Unlike conventional approaches based on population relaxation, the effect arises from coherence relaxation induced by scattering, yielding a longitudinal magnetoresistance governed by band geometry and Berry curvature. A key consequence is an unconventional scaling of the conductivity with disorder, set by the decoherence time and directly proportional to impurity density, in sharp contrast to the standard Drude expectation. This framework establishes magnetotransport as a sensitive probe of quantum coherence dynamics, offering an experimentally accessible route to quantify decoherence in solid-state systems. Moreover, the competition between external magnetic and exchange fields produces rich behavior, including a temperature-driven crossover from positive to negative magnetoresistance and a nonmonotonic temperature dependence with a conductivity maximum reminiscent of Kondo effect. The decoherence-induced magnetoresistance is a universal phenomenon, broadly present in Berry-curvature–dominated systems, including topological insulators, transition metal dichalcogenides, and spin–orbit–coupled semiconductors.

\textit{Acknowledgement}--
This work is supported by National Key R$\&$D Program of China (Grant Nos. 2020YFA0308800, 2021YFA1401500), the Hong Kong Research Grants Council (Grant Nos. 16306220, 16304523), the National Natural Science Foundation of China (Grant Nos. 12234003, 12321004, 12022416, 12475015, 11875108), the National Council for Scientific and Technological Development (Grant No. 301595/2022-4) and the São Paulo Research Foundation (Grant No. 2020/00841-9).

\end{document}